\documentclass[twocolumn,a4]{revtex4-1}
\usepackage{relsize}
\usepackage[countmax]{subfloat}
\usepackage{graphicx}
\usepackage{amssymb, amsmath,amssymb,amsfonts,dsfont}
\usepackage{amsthm,mathrsfs,amsopn}
\usepackage{dcolumn}
\usepackage{bm}
\usepackage{color}
\usepackage[utf8]{inputenc}
\usepackage[table]{xcolor}
\usepackage{floatrow}
\usepackage{lettrine}

\begin{document}

\title{{\Large Indetermination of networks structure from the dynamics perspective}}

\author{Malbor Asllani$^1$\footnote{malbor.asllani@ul.ie}, Bruno Requi\~{a}o da Cunha$^{1,2}$, Ernesto Estrada$^{3,4}$, James P. Gleeson$^1$ \vspace*{.25cm}}
\affiliation{$^1$MACSI, Department of Mathematics \& Statistics, University of Limerick,
V94 T9PX  Limerick, Ireland}
\affiliation{$^2$Rio Grande do Sul Superintendency, Brazilian Federal Police, Av. Ipiranga 1365, 90160-093 Porto Alegre, RS, Brazil}
\affiliation{$^3$Institute of Mathematics and Applications (IUMA), Universidad de Zaragoza, Pedro Cerbuna 12, E-50009 Zaragoza, Spain}
\affiliation{$^4$ARAID Foundation, Government of Aragón, Zaragoza 50018, Spain}

\begin{abstract}
Networks are universally considered as complex structures of interactions of large multi-component systems. In order to determine the role that each node has inside a complex network, several centrality measures have been developed. Such topological features are also important for their role in the dynamical processes occurring in networked systems. In this paper, we argue that the dynamical activity of the nodes may strongly reshape their relevance inside the network making centrality measures in many cases misleading. We show that when the dynamics taking place at the local level of the node is slower than the global one between the nodes, then the system may lose track of the structural features. On the contrary, when that ratio is reversed only global properties such as the shortest distances can be recovered. From the perspective of networks inference, this constitutes an uncertainty principle, in the sense that it limits the extraction of multi-resolution information about the structure,  particularly in the presence of noise. For illustration purposes, we show that for networks with different time-scale structures such as strong modularity, the existence of fast global dynamics can imply that precise inference of the community structure is impossible.
\end{abstract}

\maketitle
\noindent
Networks constitute a paradigm of complexity in real life systems by assembling the structure of the interactions of their elementary constituents~\cite{estrada_book, newman, boccaletti}. They are found at every level of biological organisation, from genes inside the cells~\cite{genetic} to the trophic relations between species in large ecosystems~\cite{trophic}. Complex networks have historically abounded in human society as well, starting with the renowed Milgram's experiment of six degrees of separation~\cite{milgram} to the impact of social media in our day~\cite{social} and the relevance of social network analysis in crime fighting~\cite{da2018topology}. In the last 20 years, science has been impacted by a huge development in the understanding of the way such complex interactions originate, looking also for universal patterns in the innumerable shapes of their structures, and investigating the consequences that such topologies of interactions have on the dynamics of the systems defined on top of complex networks.
Nowadays, with the enormous development of data science, there is a huge interest related to the network inference, namely detecting the interacting structure from external measurements or observations. For example, reconstructing the structure of brain networks from the activity of neuronal patches has been a major goal in computational neuroscience~\cite{sporns}. The dynamics that takes place on networked systems can, in some cases, strongly influence the perception that we have regarding local topological features such as the degree~\cite{malborPRL} or global ones such as  network non-normality~\cite{malborNN}. Recently, it has been shown that control methods which are based on the structural properties of networks~\cite{control} are insufficient to correctly affect the behavior of the system in the absence of insight at the dynamical level~\cite{rocha}.

In this Letter we focus specifically on the problem of measuring network centralities from the dynamical point of view. We show that the inference of networks' structural properties depends heavily on the competition between the node-based dynamics on one hand and the interactions between the nodes on the other. In particular, we illustrate such a phenomenon based on the communicability centrality~\cite{estrada, estrada_review}, considered as a reliable measure for dynamical inference~\cite{gilson}. We show that when the local intra-nodes dynamics is slower than the inter-nodes one then the ranking of the nodes according to the standard definition of the communicability, becomes inadequate. Such ranking can be enhanced if further information regarding the nature of the dynamics occurring on the network is available. Nevertheless, such corrections are based on a linear operator
~\cite{estrada_book, newman}, and the dynamical observables can shift such measures far from the topological ones in a strongly nonlinear regime. The failure of network centralities for the static case (when only the structure is considered without any consideration about the dynamical process on it) has been previously studied, with alternative approaches such as the HITS algorithm~\cite{hits} or the nonbacktracking matrix~\cite{NB} being suggested. In contrast, here, we focus on the influence that the observation of the dynamical variables in different regimes of parameters has on the distinguishability of the nodes from each other.

We start by considering a general formulation of a dynamical process in a networked system~\cite{james_book, newman}:
\begin{equation}
\dot{\textbf{x}}_i=\alpha\textbf{f}(\textbf{x}_i) + (1-\alpha)\sum_j \mathcal{A}_{ij}\textbf{g}(\textbf{x}_i,\textbf{x}_j),\, \; \forall i
\label{eq:main}
\end{equation}
where $\textbf{x}_i$ represents the multivariable vector of the state of node $i$ and $\textbf{f}(\cdot)$, $\textbf{g}(\cdot,\cdot)$ the respective intra-node and inter-nodes dynamics. The interactions are given by the adjacency matrix $\boldsymbol{\mathcal{A}}$ whose entries are $\mathcal{A}_{ij}=\mathcal{A}_{ji}=1$ if there is a undirected link going from node $j$ to node $i$ and $0$ otherwise~\footnote{The extension to strongly directed networks is done without loss of generality.}. Notice also that in order to parametrise the two effects in the dynamical system we have introduced the coefficient $\alpha$, which can be tuned to  control which part of the dynamics is more relevant.

To illustrate our analysis we will consider the $SI$ model for epidemic spreading in a metapopulation network~\cite{murray,brockmann}. {This analysis can be considered in the more general framework of metaplexes~\cite{metaplex}, where the interior of the nodes can be either a continuous or a discrete space.} Such a formulation of the spreading processes has been employed to model, for example, the propagation of  misfolded proteins in neurodegenerative diseases~\cite{proteins,neurodegener}. We assume that inside any node (cell) $i$ we have that the susceptible (regular proteins) and infected individuals (misfolding protein) will interact respectively according to $\mathcal{S}_i+\mathcal{I}_i \xrightarrow{r} 2\mathcal{I}_i$ where $r$ is the infection rate and each individual of each species will migrate between nodes $\mathcal{S}_i\xrightarrow{D} \mathcal{S}_j$ and $\mathcal{I}_i\xrightarrow{D} \mathcal{I}_j$ with a diffusion constant $D$~\cite{murray,brockmann}. The mean-field dynamics then reads:
\begin{eqnarray}
\dot{S}_i &=& -rS_iI_i + D\sum_j \mathcal{L}_{ij}S_j\nonumber\\
\dot{I}_i &=& rS_iI_i + D\sum_j \mathcal{L}_{ij}I_j\,,
\label{eq:model}
\end{eqnarray}
where $S$, $I$ are now the concentrations, respectively, of the susceptible and the infected individuals and $\boldsymbol{\mathcal{L}}$ is the Laplacian matrix defined as $\mathcal{L}_{ij}=\mathcal{A}_{ij}-k_i$ where $k_i$ is the degree of node $i$~\cite{newman}. To be compatible with the notation of eq.~(\ref{eq:main}) we have imposed $r=\alpha$ and $r+D=1$. Starting from this model, we will compare the effectiveness of measuring the nodes' centrality from the dynamical observables and compare it to different communicability definitions. To do so we first select the most central node of the graph (e.g., the one with the highest betweenness) as the observation node and then take the time needed for the infection to reach such node as the dynamical observable. More precisely, we initiate our system by infecting a single node of the graph in turn and then we find the time needed for the infection to reach a given level of concentration $I_0$ on the observation node of the network. We will indicate the observable as $RT_i$ and will refer to it as the corresponding \textit{reaching time} for the starting node $i$. Note that a similar formulation has been introduced in Ref.~\cite{brockmann} with the aim of inferring the shortest distance structure from the spreading dynamics. Here however, we will show that such inference is not reliably possible in general. It is also important to emphasise that the threshold $I_0$ is in fact a realistic consideration commonly known as the tolerance of the measurement instrument. In our case this means that it would not be possible to distinguish two nodes that have a difference in their infection level smaller than $I_0$.

The reaching time $RT_i$ for each node will be compared to the inverse of the communicability between the observable node and the one where the infection initially originated. {The reason for choosing the communicability measure as a representative of network centralities is due to the fact that it acts as a upper bound of the $SI$ dynamics (see the Supplemental Information).} The communicability for a given couple of nodes $(i,j)$ is defined as $C_{ij} = \left(e^{\beta\boldsymbol{\mathcal{A}}}\right)_{ij}=\sum_l {\left(\beta^l\boldsymbol{\mathcal{A}}^l\right)_{ij}}\big{/}{l!}$ where $\beta$ is the inverse of the temperature following the Green's function formalism~\cite{estrada_review}. This way in addition to the geodesic paths for calculating the centrality of node $i$ from node $j$, longer paths also contribute proportionally with the inverse of their respective lengths. Communicability has been initially introduced in Ref.~\cite{estrada} (with $\beta=1$) as a necessity to solve different disadvantages presented by other centrality measures based on the idea of the shortest path (betweenness, closeness etc)~\cite{estrada_book,newman}, and has since found many important applications~\cite{estrada_book, estrada, estrada_review}. An intuitive interpretation of the meaning of the communicability can be understood by considering the solution of the linear differential equation $\dot{\textbf{x}}=\boldsymbol{\mathcal{A}}\textbf{x}$. In fact, for a given node we have $x_i(t=\beta)=\left(e^{\beta\boldsymbol{\mathcal{A}}}x^0\right)_i=\sum_j C_{ij}$ where without loss of generality, we considered here $\textbf{x}^0=(1,1,\dots,1)_N$. This way the sum over all nodes of the communicability, known as the total communicability~\cite{estrada_review} (here of node $i$), is equivalent to the contribution of the flux of the system to that node once the initial condition is considered homogeneous over the network.


\begin{figure*}[t!]
\includegraphics[width=1\textwidth]{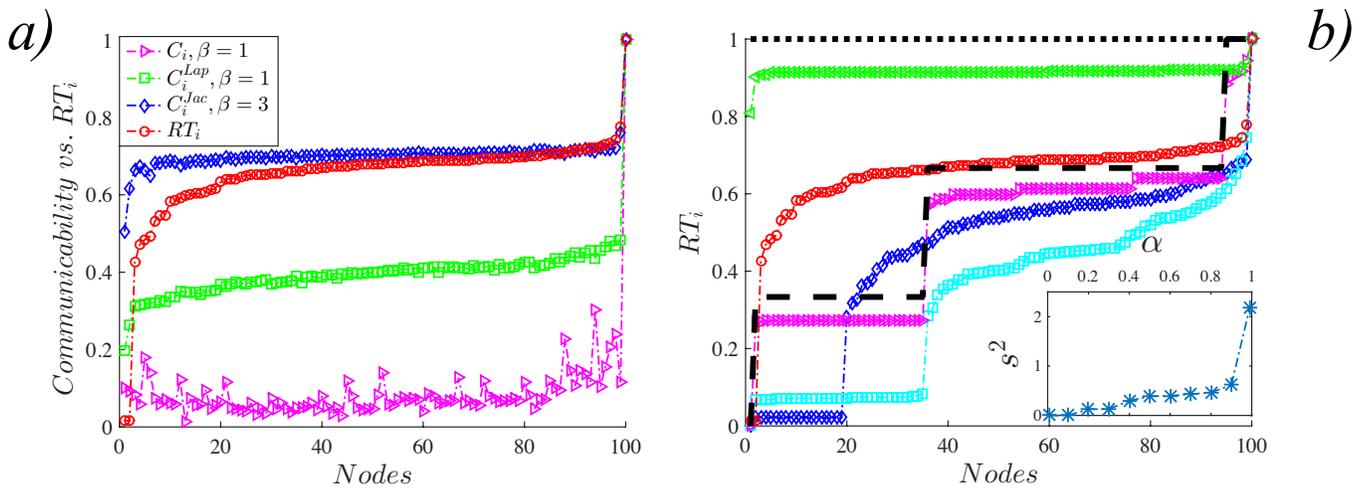}
\caption{$a)$ Comparison between the different definitions of the communicability with the reaching time $RT$ (normalised). We have represented the different centrality measures according to the several approaches of the linear dynamics (in the legend) that have been taken in consideration for the parameter $\alpha=0.33$. $b)$ Dependence of the reaching time $RT$ on the parameter $\alpha$ (green $\alpha=0.15$, red $\alpha=0.33$, blue $\alpha=0.5$, cyan $\alpha=0.91$ and magenta $\alpha\approx 1$). Tuning the parameter $\alpha$ it is possible to emphasize more one part of the dynamics than the other. In fact, in the limit when $\alpha\rightarrow 0$ all the nodes would behave the same according to the asymptote (dotted horizontal black line) losing track of the spatial structure. On the other hand when $\alpha\rightarrow 1$ we have that the $RT$ would converge at a step function (dashed black curve) which represents the distances shell from the original source of infection similarly to a contact process. In the inset is shown the variance $s^2$ as a function of the parameter $\alpha$. In both panels we used a scale-free network of $100$ nodes generated by the Barab\'asi-Albert model where also links are introduced at random with $p=0.05$ (to eventually introduce loops), and the infectiveness threshold is fixed at $I_0=0.001$.}
\label{fig:comm_comp}
\end{figure*}

Based on this interpretation, it is possible to generalise the communicability to $\tilde{C}_{ij}=\left(e^{\beta\textbf{F}(\alpha, \boldsymbol{\mathcal{A}})}\right)_{ij}$ where now $\textbf{F}(\alpha, \boldsymbol{\mathcal{A}})$ is the nonlinear operator acting on the vector of the state $\textbf{x}$ that represents the r.h.s. of eq. (\ref{eq:main}). Although such an approach should exactly capture the global dynamics, it is  of limited utility since, in general, it would require the exact knowledge of the orbits (by numerical integration) of the system which is not feasible in real scenarios. For this reason we will propose several modifications of the original communicability measure depending on the level of insight that we may have regarding the nature of the process occurring on the network. In the first attempt, we simply substitute the adjacency matrix on the exponential function of the communicability by the Laplacian matrix $\boldsymbol{\mathcal{A}}\rightarrow\boldsymbol{\mathcal{L}}$, so that $C_{ij}^{Lap} = \left(e^{\beta\boldsymbol{\mathcal{L}}}\right)_{ij}$. This can be considered a good first approximation since diffusion is a common process in networked systems~\cite{newman,james_book}. However, as we will show in the following, the Laplacian alone is often not sufficient, so for this reason we have extended the idea to the Jacobian matrix of the linearised system (around the starting steady state) $C_{ij}^{Jac} = \left(e^{\beta\boldsymbol{\mathcal{J}}}\right)_{ij}$. Note that a weak version of this has been recently proposed in~\cite{gilson} to understand the global dynamics of neuronal networks. Based on these ideas, in Fig.~\ref{fig:comm_comp} $a)$ we have first analysed the effectiveness of different definitions of the communicability versus the reaching time observable. It is clear that as one gets more insight about the nature of process and also its parameters (in this case $\alpha$), the communicability centralities appear to be more useful in understanding the dynamical observation. However, as can be noticed from the Jacobian communicability, since the dynamical observable is strongly influenced by the nonlinear terms it will not perfectly match with its corresponding communicability measure. Next, we have further investigated the competition between the spatial interactions and the internal dynamics of the nodes by tuning the parameter $\alpha$. In Fig.~\ref{fig:comm_comp} $b)$ is shown that when we change the ratio between these two parts of the system we either lose totally track of any structure in our system ($\alpha\rightarrow 0$) or we still keep some topological features in terms of shortest distances but we lose the local information on the nodes ($\alpha\rightarrow 1$). To better understand this phenomenon we return again to the general communicability now written as $\tilde{C}_{ij}=\left(e^{\alpha\beta \textbf{f}(\textbf{x})}e^{(1-\alpha)\beta g(\textbf{x},\boldsymbol{\mathcal{A}})}\right)_{ij}$ where by $g(\textbf{x},\boldsymbol{\mathcal{A}})$ we denote the operator of the flux due to the second member of the r.h.s. of eq. (\ref{eq:main}). Now it is clear that once $\alpha$ decreases the diffusion part will dominate over the interaction between the susceptible and infected individuals, so $\lim_{t\rightarrow\infty}\tilde{C}_{ij}^{\alpha\rightarrow 0}=\left(e^{\beta\textbf{f}(\textbf{x})}\right)_{ij}$. In other words, the system will first, quickly, converge to the asymptotic state of the diffusion operator (which in this case is the homogeneous fixed point), spreading in this way the seed of infection equally in each node. After the system reaches this state it is just a matter of time before all the nodes will reach the level of desired infection $I_0$  (almost) simultaneously. This explains why the nodes are not distinguishable anymore, manifested by the flatness of the reaching time vector $RT$. The random walk Laplacian~\cite{newman} is more robust to the indistinguishability in the ranking of the nodes, but using it does not qualitatively change the overall result (see the Supplemental Information). On the contrary when the parameter $\alpha$ increases we have that $\lim_{t\rightarrow\infty}\tilde{C}_{ij}^{\alpha\rightarrow 1}=\left(e^{\beta g(\textbf{x},\boldsymbol{\mathcal{A}})}\right)_{ij}$ meaning that the node where the infection is originally seeded would be (almost) immediately fully infected
while the diffusion is relatively inactive. In this case what matters is the graph distance
from the node where the infection originates. Since now the internal dynamics of the nodes is negligible the overall dynamics transforms from that of a metapopulation system to the one of a contact process~\cite{liggett}. In this extremal case the network can be considered as being organised in shells and the nodes inside each shell are not distinguishable anymore. This way we show that the possibility of inferring the shortest distance from the epidemics spreading is quite limited in a general setting, contrary to what has been argued in Ref.~\cite{brockmann}. To systematically measure the distinguishability of the nodes from each other we have also studied the sample variance $s^2=\sum_{i=1}^N (X_i-\mu)^2/(N-1)$ where $X_i$ is the sample and $\mu$ the mean value. As shown in the inset of Fig.~\ref{fig:comm_comp} $b)$ the variance increases with the parameter $\alpha$.


\begin{figure*}[t!]
\includegraphics[width=\textwidth]{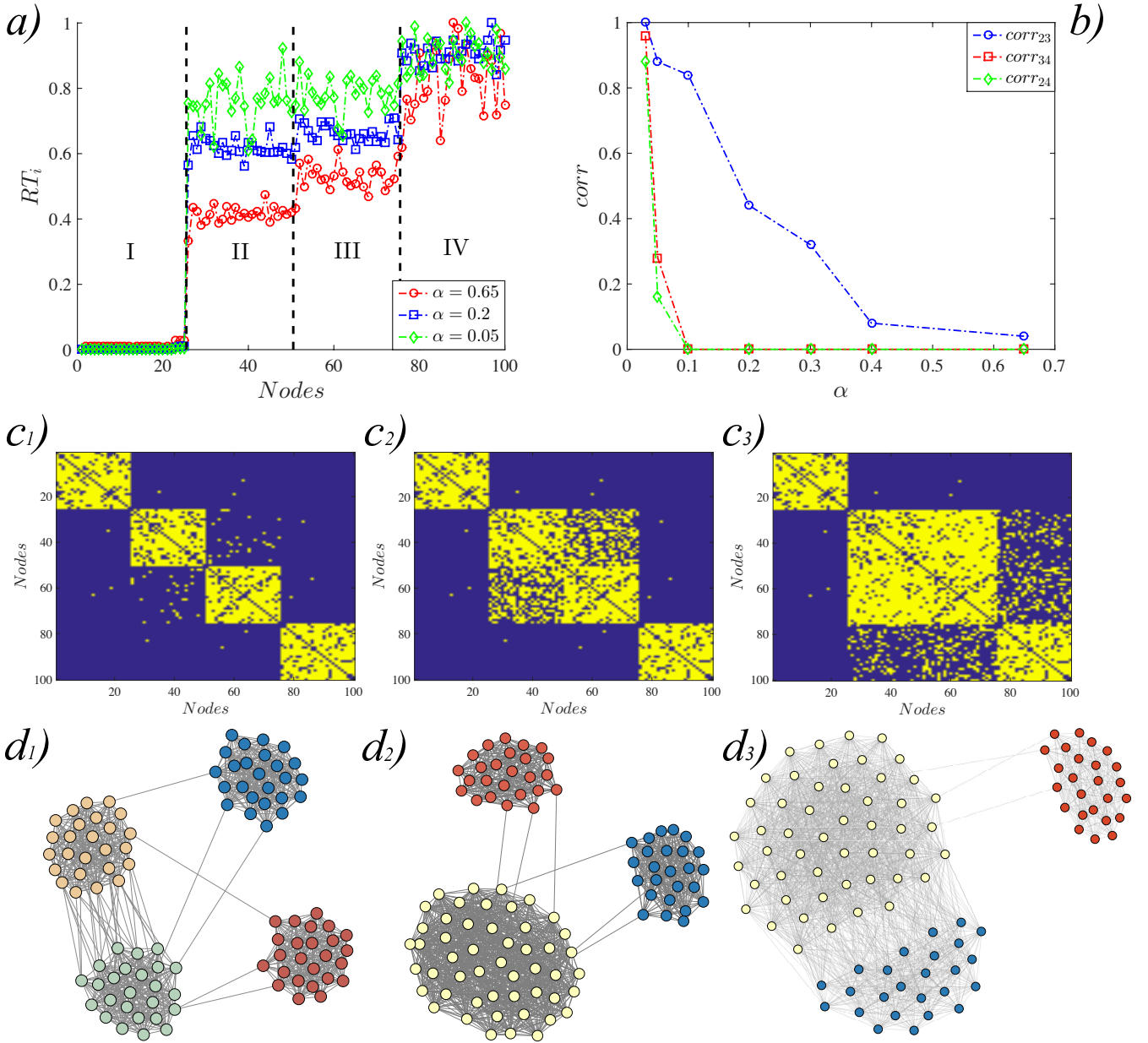}
\caption{$a)$ We plot the normalised reaching time $RT_i$ variable of the four modules (indicated by roman numbers) showing that for decreasing values of the $\alpha$ parameter (as in the legend) the ranges of the dynamical variables for different modules overlap. $(b)$ The correlation variable for each couple of modules (with the exception of the first) as a function of $\alpha$. $c)$ We show how the resolution of a given reconstruction method can be affected by different choices of the tuning parameter (for the same values as in panel $a)$). $d)$ A representative visualisation of the networks reconstruction where it is shown the gradual deformation perceived in the network modularity from:  $d_1)$ the original $4$ modules topology, $d_2$ modules $II$ and $III$ have merged and $d_3$ where module $IV$ is now merging with the union of the modules $II-III$. The modular network has $100$ nodes and has been generated through a Stochastic Block Model with total link density $p=0.2$ and
probability $0.01$ of random edge being an inter-module link.}
\label{fig:uncertain}
\end{figure*}

Heretofore we have pointed out that the importance of the difference in the time scales of the two types of dynamics that usually characterize the behavior of a networked system -- the dynamics of the node and the dynamics between the nodes-- showing that the reliability of the centrality measures decreases as the parameter $\alpha$ increases. Moreover, if the range of values taken by the reaching time $RT_i$ over all nodes $i$ is small, then in the presence of noise in the experimental data (due to the stochastic nature of the process and measurements) it is not possible to distinguish the nodes anymore. To further emphasize this point, we consider a strongly modular topology~\cite{newman,modular}. The question of inferring the modularity of real networks from data observations is of a crucial importance in  modern computational neuroscience where understanding how and why neurons or neuronal patches are organised in different communities is considered a major step forward in the comprehension of how the brain works~\cite{sporns}. It is well known that  in modular networks there are (at least) two different time-scales embedded in the structure, that of the faster intra-links (connections inside the module) compared to the slower inter-links (connections between different modules) one~\cite{multi}.
To illustrate such behaviour we will complement eq. (\ref{eq:main}) with a noise term, yielding the Langevin dynamics $\dot{\boldsymbol{\xi}}_i = \textbf{F}(\alpha, \boldsymbol{\mathcal{A}})\boldsymbol{\xi}_i + \boldsymbol{\eta}_i$ where now $\boldsymbol{\xi}_i$ is the stochastic state vector and $\boldsymbol{\eta}_i$ is the noise term with mean $\langle \boldsymbol{\eta} \rangle=0$ and variance $\langle  \boldsymbol{\eta}_i(t)  \boldsymbol{\eta}_j(t') \rangle=\sigma\boldsymbol{\delta}_{ij}(t-t')$ for each node $i$. The magnitude of the noise is given by the parameter $\sigma$. In Fig.~\ref{fig:uncertain} $a)$, we plot the reaching time variable $RT_i$ for different values of $\alpha$. Despite the presence of noise, for large value of $\alpha$ the dynamical observable is characterised by different levels or ``time slots'' for each module, so making  the modules clearly distinguishable from each other. However, as $\alpha$ gets smaller the differences between the nodes belonging to different modules decrease too,
leading to a loss of distinguishability between the modules
as shown in panel $b)$. We quantitatively estimate the correlation between different modules where for two different modules $x,y$ we have $corr_{xy}=\sum_{i=1}^M 1/M \vert \{ \min{RT^y} < RT_i^x < \max{RT^y} \}$~\footnote{Notice that $corr_{xy}\neq corr_{yx}$ in general since this definition tells how much the module $x$ belongs to the module $y$ but not necessarily the opposite is true.}. The consequences of this phenomenon are straightforward for a given inference procedure. In Fig.~\ref{fig:uncertain} in the multiple panels $c)$ and $d)$ we consider the scenario of the eventual misinterpretation of the results during the implementation of a hypothetical network reconstruction method. In fact, if the different time slots, representing the different modules, overlap with each other then it is not possible anymore for the nodes in the overlapped region, to establish to which community they belong to. In the reconstruction protocol (Fig.~\ref{fig:uncertain} $d)$) we have decided to randomly add links between two given original communities, with the number of links between modules being proportional to the correlation between their time slots for visualization purposes. As shown in Fig.~\ref{fig:uncertain} $d)$,  once $\alpha$ gets smaller all the modules (except the seeded module) appear to merge together and become indistinguishable following the previous analytical prediction.

To summarise, in this paper we have studied the question of centrality measures and consequently that of network inference in the regime of a competition between the dynamics occurring at the node level versus that at the network level. Based on the communicability measure and in a $SI$ spreading process, we have shown that, in principle, there is not a universal way to measure how central a node is related to the others unless a better understanding of the local and global dynamics is achieved. However even in this case centralities based on linear operators will fail to distinguish dynamical variables in a strongly nonlinear regime. In this sense, our results constitute an \textit{uncertainty principle} where inferring the structural properties of a network at global level means sacrificing resolution of the local dynamics of the nodes, and vice-versa. For illustration purpose, we have also considered the problem of inference of the interaction structure in a modular network. In this case the perception of modularity can be severely affected by the fast relaxation of the network global dynamics which overtakes the local one of nodes. We believe that such results can open up new scenarios of investigation in fields such as data science or computational neuroscience where  inference methods are crucial in the comprehension of the role that the different topologies of interaction have on the outcome of the dynamical process involved.

Acknowledgements: This work is partly funded by Science Foundation Ireland (grant numbers 16/IA/4470, 16/RC/3918, 12/RC/2289 P2) and co-funded under the European Regional Development Fund.

\end{document}


\title{{\Large Supplemental Information} \vspace*{.5cm}\\Indetermination of networks structure from the dynamics perspective}

\author{Malbor Asllani$^1$\footnote{malbor.asllani@ul.ie}, Bruno Requi\~{a}o da Cunha$^{1,2}$, Ernesto Estrada$^{3,4}$, James P. Gleeson$^1$ \vspace*{.25cm}}
\affiliation{$^1$MACSI, Department of Mathematics \& Statistics, University of Limerick,
V94 T9PX  Limerick, Ireland}
\affiliation{$^2$Rio Grande do Sul Superintendency, Brazilian Federal Police, Av. Ipiranga 1365, 90160-093 Porto Alegre, RS, Brazil}
\affiliation{$^3$Institute of Mathematics and Applications (IUMA), Universidad de Zaragoza, Pedro Cerbuna 12, E-50009 Zaragoza, Spain}
\affiliation{$^4$ARAID Foundation, Government of Aragón, Zaragoza 50018, Spain}

\maketitle
\subsection*{Communicability as an upper-bound for the SI dynamics}

\noindent
In the main text we used the communicability measure as a representative of the network centralities. Here we will illustrate the reason behind this choice. In fact, it is known that the linearization of the $SI$ model for the infected species around the respective initial fixed point $I_i^*=0, \forall i$  is given by

\begin{equation}
\dot{\textbf{I}}(t)=\gamma \boldsymbol{\mathcal{A}}\,\textbf{I}(t)\label{eq:model}
\end{equation}
and it is exponentially unstable. In particular, the linearized problem comes from the observation that

\begin{equation}
\dot{I}_{i}(t)=r[1-I_{i}(t)]\sum_{j=1}^{N}\mathcal{A}_{ij}I_{i}(t)\leq r\sum_{j=1}^{N}\mathcal{A}_{ij}I_{i}(t)
\end{equation}
or

\begin{equation}
\dot{\textbf{I}}(t)\leq r \boldsymbol{\mathcal{A}}\,\textbf{I}(t),
\end{equation}
 $\forall i$ and $\forall t$, which means that the linear dynamical system $\dot{\textbf{I}}(t)=\gamma \boldsymbol{\mathcal{A}}\,\textbf{I}(t)$ is an upper-bound for the original non-linear dynamical system. Consequently, a solution for the linearized problem can be written as

\begin{equation}
\textbf{I}(t)=e^{rt\boldsymbol{\mathcal{A}}}\textbf{I}_{0}.
\end{equation}
It has been recently shown by Lee et al. \cite{Lee} that after an appropriate renormalisation of $I_{i}(t)$, the solution of the $SI$ model can be written uniquely (apart from constants) in terms of
$e^{\gamma rt\boldsymbol{\mathcal{A}}\text{\,}}$, where $\gamma$ is a normalisation parameter. In this context, it is then clear that the communicability function~\cite{estrada}, which is defined for a couple of nodes as $C_{ij} = \left(e^{\beta\boldsymbol{\mathcal{A}}}\right)_{ij}=\sum_l {\left(\beta^l\boldsymbol{\mathcal{A}}^l\right)_{ij}}\big{/}{l!}$, {is the most suitable choice of centrality to measure the dynamics on the network.} We should notice that $\beta$ is here a parameter that globalises the other parameters present in the solution of the $SI$ model. It can be interpreted as an inverse temperature following the Green's function formalism~\cite{estrada_review}.

\subsection*{The case of the random walk Laplacian}

\begin{figure}[h!]
\includegraphics[width=\textwidth]{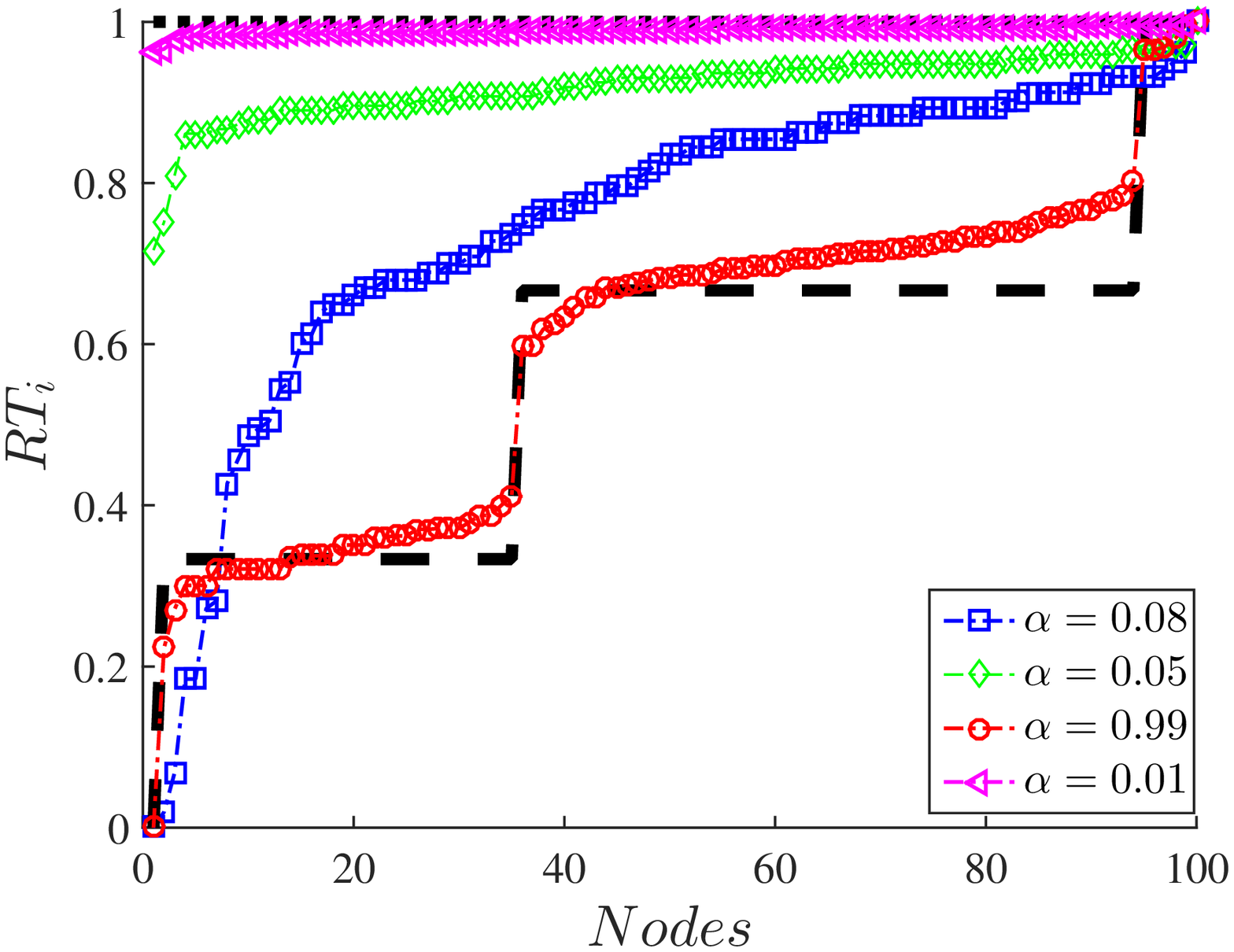}
\caption{Different reaching time vector $RT$ curves for the random walk Laplacian formulation. The different values of the parameter $\alpha$ can be found in the figure legend. The dotted and dashed black lines represent again the asymptotic behavior as in Fig. 1 $b)$. The initial conditions are $x_i(0)=0.01$.}
\label{fig:Fig3}
\end{figure}

\noindent
In this section we will briefly discuss the scenario when other cases beside the combinatorial Laplacian operator used thorough the main text, are considered for modelling the diffusion process. A well known operator used extensively in network science for describing the collective dispersion dynamics of group of individuals is the random walk Laplacian $\boldsymbol{\mathcal{L}}^{RW}$ matrix whose entries are defines as  $\mathcal{L}^{RW}_{ij}=\mathcal{A}_{ij}/k_j-\delta_{ij}$~\cite{estrada_book,newman}. Random walk diffusion has been used in a broad range of applications from human mobility, to brain dynamics, to social media. An important fact related to the random walk operator is that it does not relax at a homogeneous equilibrium. In fact, the steady state for each node of this operator is proportional to its degree $x_i(\infty)\propto k_i$. Nevertheless this property should not distort our vision of the qualitative behavior predicted and decribed in the main text. So following the analysis along the same lines as above we expect that for small values of the control parameter $\alpha$ the ranking of the nodes should gradually fade. However, as it can be observed by Fig.~\ref{fig:Fig3} such behavior starts to become evident only for very small values of $\alpha$.

The explanation behind this outcome should be found in the mathematical definition of the random walk Laplacian. In fact using the analogy with the combinatorial Laplacian, the random walk one can be defined as $\boldsymbol{\mathcal{L}}^{RW}=\boldsymbol{\mathcal{L}}\boldsymbol{\mathcal{K}}$ where $\boldsymbol{\mathcal{K}}=diag(1/k_1, 1/k_2,\dots,1/k_N)$. So in this sense we can think of the random walk diffusion as governed by the standard combinatorial Laplacian but where the diffusion rate $1-\alpha$ is now weighted locally by the degree of the neighbour nodes in the way described in the definition of $\boldsymbol{\mathcal{L}}^{RW}$. For this reason it is expected that the ratio of the two components of the system (the local node dynamics versus the global network one) is not simply controlled by the parameter $\alpha$ but also by the degree distribution of the network under discussion. This dependence is twofold: on one hand it depends on how dense the network is (e.g. the mean degree $\langle k\rangle$) and on the other hand it depends on the distribution of degrees in the sense that the diffusion on a central hub node would be quite different from a peripheral  leaf node. From this point of view a random walk diffusion can be considered more robust in terms of losing the ranking of the nodes in the dynamical observable, the reaching time $RT$ in our case. In general, a network process whose dynamics depends implicitely on the local structural properties may behave differently from the explicit parametrisation of its importance inside the system dynamics. Nevertheless this does not change the qualitative outcome discussed in the main text that no distinguishability of the nodes is possible for fast network dynamics compared to the slow nodes' one.

\